\documentclass[prd,showpacs,preprint,amsmath,amssymb,nofootinbib]{article}
\usepackage{amsmath,amssymb,graphics,epsfig}
\usepackage{enumerate}
\usepackage{theorem}

\begin{document}

\title{\bf Cosmic magnetization in curved and Lorentz violating space-times}

 \author{Alexandros P. Kouretsis\\ {\small Section of Astrophysics, Astronomy and Mechanics, Department of Physics}\\ {\small Aristotle University of Thessaloniki, Thessaloniki 54124, Greece}}

\date{}

\maketitle

\begin{abstract}
  The presence of the large-scale magnetic fields is one of the greatest  puzzles of contemporary cosmology. The symmetries of the electromagnetic field theory combined with the geometric structure of the FRW universe leads to an adiabatic decay of the primordial magnetic fields. Due to this rapid decay the residual large scale magnetic field is astrophysically unimportant. A common feature among many of the proposed amplification mechanisms is the violation of Lorentz symmetries. We introduce an amplification mechanism within a Lorentz violating  environment where we use Finsler geometry as our theoretical background. The mechanism is based on the adoption of a local anisotropic structure that leads to modifications on the Ricci identities. Thus, the wave-like equation of any vector source, including the magnetic field, is enriched by the Finslerian curvature theory. In particular limits the remaining seed field can be strong enough to seed the galactic dynamo. In our analysis we also develop the 1+3 covariant formalism for the 4-vector potential in curved space-times.
\end{abstract}

\section{Introduction}

Astrophysical observations indicate that magnetic fields are widespread in the universe \cite{Magevery}.  Their presence  is confirmed in almost every gravitationally bound system, ranging from stars up to  faraway  galaxies and clusters of galaxies. The typical strength of the intergalactic magnetic field in the Milky Way  and other spiral and barred galaxies is of the order of $B_{galactic}\sim 10^{-6}G$. In addition, clusters of galaxies are permeated by magnetic fields of almost the same intensity with those in galaxies despite the scale difference. The fact that gravitationally bound systems of different scale are permeated by magnetic fields of the same strength indicates that they might have common origin. Moreover, according to recent analysis of the data from Fermi and Hess telescopes, there is evidence of coherent intergalactic magnetic fields in low density regions with strength between $\sim 10^{-17}G$ and $\sim 10^{-14}G$ \cite{Intergal}. Furthermore, there are strong indications of large scale magnetic field in filaments \cite{Battaglia}.

Up to a point,  galactic magnetism can be explained as a result of local astrophysical mechanisms. These  rely on magnetohydrodynamic turbulence and on the differential rotation of galaxies. In particular the differential rotation (galactic dynamo) \cite{Dynrev} can lead to the exponential amplification of a seed field by a factor $\sim e^{\Gamma \Delta t}$. The exponential index $\Gamma$ depends on the parameters of the particular dynamo mechanism, which in the literature range between $\sim 0.45 Gyr^{-1}$ and $\sim 5 Gyr^{-1}$ \cite{Gyr}. In addition, we expect that the seed field for the galactic dynamo has been already amplified by a factor $10^4$, due to magnetic flux conservation during protogalaxy collapse. Putting these together, in optimistic scenarios  we need a seed field $B_{seed} \gtrsim 10^{-33}G $ while in the worst case we get $B_{seed} \gtrsim 10^{-15} $ \cite{Campanelli:2007cg}. Whether such fields are of primeval origin or they have been generated during galaxy formation remains unclear. However, the presence of magnetic fields in remote protogalaxies and the lower limits retrieved in intergalactic voids suggest that the primordial hypothesis might be the case. At the same time, future observations of the cosmic microwave background (CMB) may confirm the cosmological origin of the intergalactic magnetism by reporting imprints on the CMB spectrum (for a recent review see \cite{Durrereview}).

The correlation length of magnetic fields generated during the post-inflation era is restricted by the fact that physical mechanisms must be causal. Hence, the coherence scale of the seed field (e.g. generated  in a phase transition \cite{Kahniashvili:2012uj}) must be subhorizon. Consequently, the correlation length drops well below $10kpc$ which is typically the lowest requirement of the dynamo mechanism. We can overcome this obstacle by assuming some degree of turbulence, resulting to an increase of the coherence scale by an inverse cascade process (see \cite{Widrow:2011hs} and references therein). However, this mechanism seems to require large amounts of magnetic helicity. An alternative mechanism to produce seed fields correlated in scales $10kpc\sim1Mpc$ is inflation \cite{Turner:1988}. The latter relates microphysical processes to large scale phenomena through a rapid expansion of the cosmological fluid. Precisely, during de Sitter phase, electromagnetic  quantum fluctuations are exponentially stretched and cross outside the horizon within a finite time interval. The drawback is that magnetic fields decay adiabatically throughout the whole cosmological history resulting in an astrophysically irrelevant magnetic intensity. This is a consequence of  the conformal invariance of the electromagnetic field theory   and of the conformal flatness of the Friedmann Robertson Walker (FRW) metric.

The aforementioned strength problem  is confronted by several proposed amplification mechanisms, operating within or beyond the framework of conventional electromagnetism (see for example \cite{ampmec}).  The vector nature of the electromagnetic field guarantees that it couples to the curvature of space time through the Ricci identities \cite{Tsagas:2004kv}. Generally speaking, this clearly indicates that non-trivial geometric structures directly affect the evolution of seed fields, since the commutator of covariant derivatives is directly related to the symmetries of space-time. Moreover, if we depart from the safe harbor of conventional electrodynamics there is a plethora of amplification mechanisms. Most of these break the conformal invariance of the electromagnetic action or extend the geometric structure of the physical manifold \cite{Durrereview,Widrow:2011hs}. It is well-known, that Lorentz invariance is strongly related to the symmetries of the photon sector and determines the local structure of space-time. In fact, a common feature among many of the inflationary amplification mechanisms is  Lorentz violation (LV). As noted in \cite{Kostelecky},  if Lorentz invariance is not an exact symmetry of nature we expect that conformal invariance breaks down. In addition, the Ricci identities are incompatible with many LV scenarios and thus we expect that the electromagnetic wave equation will be affected by the modified curvature theory.

It is widely believed that LV are candidate signals of an underlying unified quantum gravity theory (QG). We expect these to emerge in highly curved regions of space-time, i.e. close to the classical singularities of general relativity (GR). Thus, the primeval universe is a possible laboratory to test LV through  their imprints on the cosmological relics. In that case, if galactic magnetism is of cosmological origin it may  carry information about QG physics, since the inflationary era, where quantum fluctuations are generated, is some orders of magnitude close to the Planck scale. In fact, there are several studies on the connection between LV and the survival of cosmological magnetic fields \cite{Bertolami}. The basic framework to build a realistic effective field theory for LV in curved space-time is the Standard-Model-Extension (SME) \cite{SME}, where particles follow Finslerian geodesics \cite{Kostelecky:2012ac}. In a recent series of papers \cite{Campanelli:2008xs}, the link between SME and primordial magnetism was studied and estimations about LV parameters were given with respect to the observed intergalactic magnetic fields.

In this paper, we study the wave equation of the electromagnetic field  by dropping Lorentz invariance. Motivated by the previous analysis we use Finsler geometry as our theoretical background to study LV (see for example \cite{Rad41}). In addidion,  Finsler geometry is encountered in many branches of QG like SME \cite{Kostelecky:2012ac}, Horava-Lifshitz gravity \cite{Romero:2009qs}, holographic fluids \cite{Leigh:2011au}, D-particle space-time foam \cite{Mavromatos:2010ar}, very special relativity \cite{Cohen:2006ky}, Galilean transformations in curved spacetime \cite{Germani:2011bc}, bi-metric theories of gravity \cite{Skakala:2010hw} and plays a keynote role in the analogue gravity program \cite{Barcelo:2005fc}. In Finsler geometry the local structure of GR is extended by adopting a non-quadratic line element \cite{Shen,RundFB,Miron}. The deviation from the local symmetries of GR is measured by a pure geometric entity, the color (or equivalently non-quadraticity) \cite{Shen}. The color together with curvature determines the LV-kinematics. Due to the local anisotropy of the Finslerian manifold the curvature theory is modified in alliance with the QG phenomenology where we expect that  LV are incompatible with the Ricci and the Bianchi identities \cite{Kostelecky:2003fs}.

Central role to our analysis plays the 1+3 covariant formalism \cite{Relcos} that we can extend in Finsler geometry \cite{Kouretsis:2012ys}. Firstly, we develop the 4-vector potential representation of electromagnetism in curved space-times and we discuss the FRW inflationary scenario. This unfolds in a covariant way the effect of curvature in the wave equation of the electromagnetic field. Then, we generalise the metric structure of the effective manifold by relaxing the local symmetries of GR. As a consequence, the Ricci identities are modified and this leads to a `colored' (LV) generalisation of the wave equation for the vector potential. Under particular conditions the LV contribution can superadiabatically amplify the magnetic quantum fluctuations. Finally, we provide order of magnitude estimations for the residual magnetic intensity with respect to the scale of inflation and the `Finslerity' of space-time.

\section{Finsler geometry}\label{fin}

In a differentiable manifold $\mathcal M$ the metric structure is determined by a scalar function $F(x,y)$ on the tangent bundle $\mathcal{TM}$, where $x^a$ is the position variable  while $y^a$ is a tangent vector field and the pairs $(x^a,y^a)$ is a coordinate system of $\mathcal TM$ over a local region of $\mathcal M$.   In Riemann geometry the metric function $F(x,y)$ is  the norm of the $y^a$-vector field and gives back the length of the tangent $y$ at a point $x$. It is quadratic and homogeneous of first degree  with respect to the $y^a$ increments. The first property reflects that every tangent space is Euclidian (in relativity Minkowski) and secures the local flatness of the manifold. Therefore, it is directly related to the $SO_4$ local group of GR which implies the form of the metric function, $F(x,y)=\sqrt{n_{ab}y^ay^b}$, where $n_{ab}$ is the Minkowski metric in an infinitesimal region of the Riemannian space-time. The second property implies that the length between two points is independent of the parametrization of the connecting curve.  Finsler geometry is just Riemann geometry without the quadratic restriction on the metric function \cite{ShenAms}. The infinitesimal flat neighborhood is replaced by an anisotropic structure and this  affects the local as well as the global properties of the manifold.

The non-quadraticity is an extra geometric property named as the {\it color} \cite{Shen}. The color, the curvature and the interplay between them determine the evolution of the Finslerian congruences. Using the first order homogeneity of the metric function $F(x,y)$ we can introduce the Finsler metric tensor
\begin{equation}
g_{ab}(x,y)={1\over2}\frac{\partial^2F^2}{\partial y^a\partial y^b}.\label{mten}
\end{equation}
The $y$-dependence of the metric is a direct consequence of the non-quadraticity. Note, that Riemann geometry is a special case and it is retrieved when we impose quadraticity on the distance module. The dependence on the fiber coordinates $y^a$ directly reflects the Lorentz violating structure of the Finslerian space-time. They may be physically interpreted as an arbitrary direction at each tangent space induced by the breaking of Lorentz invariance (see for example \cite{Kouretsis:2012ys,Ishikawa}). In fact, Finsler geometry is encountered in Lorentz violating branches of quantum gravity \cite{Romero:2009qs,Leigh:2011au,Mavromatos:2010ar,Cohen:2006ky,Germani:2011bc,Skakala:2010hw,Barcelo:2005fc} and also effectively describes motion in anisotropic media \cite{Born}. Apart from the metric tensor (\ref{mten}) there is another important geometric entity, the Cartan's tensor
\begin{equation}
C_{abc}={1\over2}{\partial g_{ab}\over \partial y^c},\label{Cten}
\end{equation}
that measures departures from the quadratic measurement. In other words, the tensor field (\ref{Cten}) monitors the colorful 'morphology' of the Lorentz violating medium. When $C_{abc}=0$ the manifold is white (Riemann) and the Lorentz symmetry is restored. Noteworthy is the property of the the Cartan tensor that follows from the first order homogeneity $C_{abc}y^c=0$.

In general, Finsler is a fiber space geometry since tensor fields depend both on the position and on the $y$-increments and follow the linear coordinate transformation law
\begin{equation}
T_{ab}(x,y)={\partial \tilde x^c\over \partial x^a }{\partial \tilde x^d\over \partial x^b}\tilde T_{cd}(\tilde x,\tilde y).
\end{equation}
The pair $(x,y)$ is the element of support and introduces an arbitrary direction on each tangent space. Hence, the parallel displacement of a vector field $X^a$ along a Finslerian congruence $u_a(x,y)$ is given with respect to the  supporting direction $y^a$  \cite{RundFB}. In this setup, of particular interest are the Finslerian congruences $\gamma({\tau})$, with tangent $u_a(x,y)$, along which the absolute derivative of the supporting direction vanish, namely $D y^a/d\tau=0$. In that case, the parallel displacement of a tensor field along the restricted bundle is similar with the Riemannian limit. After imposing metricity, it is given with respect to the Cartan connection by the following formula
\begin{equation}
\dot X^{ab...}_{...cd}=X^{ab...}_{...cd|e}u^e.\label{tder}
\end{equation}
The $|$ operator is the only covariant derivative involved in the kinematics for our restrained setup. For simplicity, in the rest of our analysis  we will focus in geodesic congruences along which the parallel translation (\ref{tder}) is valid.

 Consider that the average motion of matter is given by the restricted  1rst rank tensor field $u^a$ that is a time-like normalized direction, $u^au_a=-1$. Then, the affine parameter $\tau$ is the proper time of the family of fundamental observers. Hence, the parallel translation (\ref{tder}) stands for the time derivative along the fluid flow lines. Also, we can introduce the projected covariant derivative on the instantaneous rest frame
\begin{equation}
{\rm D}_aX^{bc...}_{\;\;\;\;...de}=h^{f}_{\;\;a}h^{b}_{\;\;g}h^{c}_{\;\;h}h^{i}_{\;\;d}h^{j}_{\;\;e}X^{gh...}_{\;\;\;\;...ij|f}\label{sder}
\end{equation}
where we define the tensor $h_{ab}=g_{ab}+u_au_b$ that projects orthogonal to the observers' time-like flow. With the aid of the projection tensor $h_{ab}$ we can decompose space-time quantities to their time-like and space-like components. For example, given a 4-vector field $X^a$ its' irreducible decomposition withe respect to the $u^a$ time-like congruence is
\begin{equation}
X^a=(h^{ab}-u^au^b)X_b=-Xu^a+\mathcal X^a
\end{equation}
where $X=X^au_a$ is the time-like part and $\mathcal X^a$ is the projection on the instantaneous rest space (for further details see \cite{Relcos,Kouretsis:2012ys}). Using the time derivative (\ref{tder}) and the spatial derivative (\ref{sder}) together with the Finslerian curvature theory we can investigate the evolution of physical fields in a full covariant way. When the space-time manifold is white (Riemann), $C_{abc}=0$, the above definitions reduce to the usual GR operators.

\section{Vector potential in GR's 1+3 formalism}\label{Rcove}

An equivalent formalism to construct the electromagnetic field theory is the 4-potential representation. In the following we develop the latter case using the Ehlers\,-\,Ellis $1+3$ formalism mainely in the Coulomb gauge. We use this method to underline in a relatively straightforward way the role of Riemann curvature in the photon sector.  This will prove useful in Sec.\ref{LVwav} where we will generalise the space-time manifold to be Finslerian.
The physics of the photon sector are encrypted in the 4-vector $A_a$ that corresponds to a $U(1)$ gauge field
\begin{equation}
F_{ab}=\nabla_b A_a-\nabla_a A_b.
\end{equation}
The above tensor field remains invariant under any gauge transformation, $A_a\rightarrow A_a'+\nabla_a f$. This property is responsible for the freedom of fixing the gauge in a convenient way depending on the physical problem.

In studies of cosmological magnetic fields the most widespread gauge is the Lorentz gauge, $\nabla^aA_a=0$. The last condition is written to its irreducible parts as
\begin{equation}
\dot{\phi}+\Theta\phi-{\rm D}_a\mathcal{A}^a-\dot{u}_a\mathcal{A}^a=0,\label{lgau}
\end{equation}
where we use the definitions, $\Theta={\rm D}^au_a$ for the expansion of the cosmic medium, $\phi=A_au^a$ for the time-like part of the 4-vector potential  and $\mathcal{A}_a=h_a^{\;\;b}A_b$ for the projection on the instantenious rest frame of the fundamental observer. Nevertheless, we have not completely  exhausted the gauge freedom. Without loss of generality we can further introduce the Coulomb sub-gauge, ${\rm D}^a\mathcal{A}_a=0$. Then, the last term of the propagation relation (\ref{lgau}) for the scalar potential is the only source term. The latter  term implies a time-varying scalar potential even in the absence of expansion and reflects the fact that we treat space-time as a single entity. This is an analogous relativistic effect with the contribution of the 4-acceleration  in the Maxwell equations in the covariant 1+3 representation \cite{Barrow:2006ch}.

Consider a slightly magnetised FRW model where the energy density of the electromagnetic field is a first-order perturbation. Then, adopting the Coulomb gauge and keeping up to first order terms, relation (\ref{lgau}) is consistent with the condition, $\phi=0$. In that case, the electric field $E_a=F_{ab}u^b$, is written with respect to the 3-vector potential as
\begin{equation}
E_a=\dot{\mathcal{A}}_{\langle a\rangle}+{1\over3}\Theta\mathcal{A}_a,\label{elA}
\end{equation}
while the magnetic field $B_a=\epsilon_{abc}F^{bc}/2$ is given by
\begin{equation}
B_a=-{\rm curl}\mathcal{A}_a,\label{magA}
\end{equation}
where we define ${\rm curl}\mathcal{A}_a=\epsilon_{abc}{\rm D}^b\mathcal{A}^c$, while angle brackets stand for the projection perpendicular to the bulk flow of matter $X_{\langle a \rangle}=h_{a}^{\;\;b}X_b$.
 Introducing the conformal time ${\eta}=\int{dt\over a}$ for the scale factor of the medium, $\Theta=3{\dot a\over a}$, and using  (\ref{elA}) and (\ref{magA}) we retrieve the well known relation $B\sim k\eta E$, where $k$ is the wavenumber while $B$ and $E$ are the average magnetic and electric intensities.

The physical meaning of the 3-vector potential $\mathcal{A}_a$ is not direct. However, using the relations (\ref{elA}) and (\ref{magA}) we can translate it to the physically meaningful electric and magnetic spatial vectors. The 4-vector potential satisfies the electromagnetic field equation
\begin{equation}
\nabla^b F_{ab}=J_a.\label{elefield}
\end{equation}
In general, the time-like part of the previous relation corresponds to the wave-like equation for the scalar part of $A_a$, while the projection to the rest space of the fundamental observer gives back the wave-like equation for the 3-vector potential $\mathcal{A}_a$. In our slightly magnetised FRW background and for the Coulomb gauge with $\phi=0$, the projection of relation (\ref{elefield}) along the observers flow-lines is a trivial identity while the space-like part corresponds to the following wave-like expression
\begin{equation}
\ddot{\mathcal{A}}_{\langle a\rangle} +\Theta \dot{\mathcal{A}}_{\langle a \rangle}+{1\over3} \left(\dot{\Theta}+\frac{2}{3}\Theta^2\right) \mathcal{A}_a-
{\rm D}^2\mathcal{A}_a+{\rm D}^b{\rm D}_a\mathcal{A}_b=-\mathcal{J}_a,\label{wavA}
\end{equation}
where $\mathcal{J}_a=h_{a}^{\;\;b}J_b$ is the spatial current. The last term in the lhs of (\ref{wavA}) is directly related to the intrinsic curvature of the instantaneous rest space. In particular, for vanishing vorticity the instantenious space is a Riemannian hypersurface and the induced covariant derivative commutes in the usual way for any 3-vector
\begin{equation}
2{\rm D}_{[c}{\rm D}_{b]}\mathcal{A}_a=\mathcal{R}_{dabc}\mathcal{A}^d\label{tric}
\end{equation}
where $\mathcal{R}_{dabc}$ is the intrinsic curvature of the spatial slice. The intrinsic 3-curvature $\mathcal{R}_{dabc}$ is related to the space-time curvature by projecting the 4-Ricci identities on the spatial hypersurfaces. Precisely, the spatial curvature is given with respect to the projected 4-curvature through the Gauss-Godazzi formula
\begin{equation}
\mathcal{R}_{dabc}=h_{a}^{\;\;q}h_{b}^{\;\;s}h_{c}^{\;\;f}h_{d}^{\;\;p}R_{qsfp}-\nu_{ac}\nu_{bd}+\nu_{ad}\nu_{bc},\label{gaugo}
\end{equation}
where $\nu_{ab}={\rm D}_b u_a $ is the extrinsic curvature of the spatial hypersurface and coincides with the relative flow tensor of neighboring observers.

Using the three Ricci identity (\ref{tric}) and taking into account the Coulomb gauge and the Ohm's law $\mathcal{J}_a=\sigma E_a$, the wave-like relation (\ref{wavA}) is reexpressed in conformal time  as
\begin{equation}
 \tilde{\mathcal{A}}_a''+a\sigma\tilde{\mathcal{A}}_a'-a^2{\rm D}^2\tilde{\mathcal{A}}_a+a^2\mathcal{R}_{ba}\tilde{\mathcal{A}}^b=0\label{cwav}
\end{equation}
where we define the rescaled 3-vector potential $\tilde{\mathcal{A}}_a=a\mathcal{A}_a$, while the prime denotes differentiation with respect to the conformal time $\eta$. The conductivity acts as a friction term while the effect of spatial curvature depends on the type of the 3-geometry. In the FRW case, the  spatial sections are hypersurfaces of constant curvature
\begin{equation}
 \mathcal{R}_{abcd}={K\over a^2}(h_{ac}h_{bd}-h_{ad}h_{bc}),\;\;\;\;\;\;K=0,\pm1
 \end{equation}
and relation (\ref{cwav}) is written for the k-th mode as
\begin{equation}
\tilde{\mathcal{A}}_{(k)}''+a\sigma\tilde{\mathcal{A}}_{(k)}'+(k^2+2K)\tilde{\mathcal{A}}_{(k)}=0.\label{kwav}
\end{equation}
In the lhs of the above relation, when the conductivity is high the friction term dominates over the last term. The high conductivity limit implies $\tilde{\mathcal{A}}_{(k)}'\rightarrow0$ which translates through relation (\ref{elA}) to a vanishing electric field, while relation (\ref{magA}) gives back the adiabatic decay for the mean magnetic field, $B\propto a^{-2}$.  Also, during inflation where the conductivity is low we always retrieve an adiabatic decay for the magnetic energy density \footnote{We can recast the covariant expressions (\ref{elA}) and (\ref{magA}) with respect to the conformal time and the rescaled 3-vector potential as $a^2E_a=\tilde{ \mathcal{A}}_a'$  for the electric field and $a^2B_a=-a{\rm curl}\tilde{\mathcal{A}}_a$. Thus, for the mean values of the electromagnetic field we retrieve the qualitative relations $a^2E\propto \tilde{\mathcal{A}}/\eta$ and $a^2B\propto \tilde{\mathcal{A}} $. }. The low conductivity is a good approximation within an inflationary era since causally connected regions grow rapidly beyond the horizon and the temperature rapidly decays. On the other hand, the cosmological medium is a highly conducting fluid in the post inflation era since $\sigma\sim T/\alpha$, where $\alpha$ is the fine structure constant. Nevertheless, one can argue that in the post inflation era spatial currents are  not important for superhorizon modes since the latter belong to acausal regions and we may omit the conductivity term in relation (\ref{cwav}). We point out that in the literature there are contradicting examples about the contribution of spatial currents in the wave equation of superhorizon modes (see for example \cite{Widrow:2011hs,Subramanian:2009fu}).

In case of a flat FRW universe (for the open FRW case see \cite{openamp}, but see also \cite{Adamek:2011hi}) the Hubble horizon, $\lambda_H=H^{-1}$, defines the causally connected regions. Moreover, in de Sitter inflation the Friedmann equation of motion implies
$H^2\sim M^4/m_{pl}^2$, where $M$ is a constant that stands for the scale of inflation and $m_{pl}$ is the Planck mass. In this era, the exponential growth of the scale factor secures a constant Hubble horizon. On the other hand, the physical wave-length of a quantum fluctuation, $\lambda_{phys}=a\lambda$, generated in the de Sitter background grows exponentially. Therefore, the  length of the  perturbation becomes superhorizon in  finite time.  Hence, if inflation continues for some e-folds after first horizon crossing the coherence of the seed field can be today at scales of $10kpc-Mpc$. This mechanism solves the problem of the coherence but the exponential expansion combined with the adiabatic decay of $\rho_B$ leads to astrophysically unimportant magnetic fields.

In order to estimate the energy density of the survived magnetic field we need to impose some initial conditions. Let us consider the starting value of the optimistic starting value of the  seed field at the first horizon crossing, namely $\lambda_H=\lambda_{phys}$, as $\left(\rho_B\right)_{HC}\sim H^4 $. Furthermore, for instantaneous reheating ($M\sim T_{RH}$), the fraction of the scale factor from the first horizon crossing until  the end of inflation is approximately
\begin{equation}
\frac{a_{INF}}{a_{HC}}=e^N\sim 10^{26}\frac{\lambda_0}{Mpc}\;\frac{M}{m_{pl}},
\end{equation}
where $\lambda_0$ is the observed length of the generated quantum fluctuation and $N$ stands for the e-folds. Also, from the end of inflation until the present epoch due to entropy conservation we get for the scale factor $a_0/a_{INF}\sim10^{29}(H/10^{-5}m_{pl})^{1/2}$. Evolving adiabatically  the magnetic energy density for the initial condition $\left(\rho_B\right)_{HC}$, the residual field is
\begin{equation}
B_0\sim10^{-58}\left(\lambda_0\over Mpc\right)^{-2}G.\label{magin}
\end{equation}
As it is clear from the above relation the resulting  magnetic intensity is independent of the inflation's scale. However, the today values of the magnetic field (\ref{magin}) for coherent scales $10kpc-Mpc$ are extremely weak compared to the observations. They are also astrophysically irrelevant  even for very optimistic dynamo mechanisms where a seed field of the order $B_{seed}\sim10^{-30}G$ is required. This problem inspires the study of modified electrodynamics within an inflationary scenario to slow down the dilution of the magnetic intensity.

\section{The LV wave equation}\label{LVwav}

Consider the $U(1)$ gauge field within a Finslerian space-time. Then, for the particular geometric set-up that we discussed in section (\ref{fin}) the electromagnetic 2nd rank tensor
is written as
\begin{equation}
F_{ab}=A_{a|b}-A_{b|a}\label{Ften}
\end{equation}
for the Finslerian covariant derivative (\ref{tder}). For the sake of simplicity, we will investigate the particular generalization of the Riemannian case, $A_a=A_a(x)$. In this set up, it is trivial to prove that the tensor field (\ref{Ften}) is invariant under the gauge transformation $A_a\rightarrow A'_a+f_{|a}$. Thus, following the same arguments with the standard relativistic electrodynamics (see Sec.\ref{Rcove}) we can introduce the Coulomb gauge in a covariant way, namely
\begin{equation}
\phi=A_au^a=0\;\;\;\;,\;\;\;\;{\rm D}_a \mathcal{A}^a=0.
\end{equation}
In that case, the electric field $E_a=F_{ab}u^b$ and the magnetic field $B_a=\epsilon_{abc}F^{bc}/2$ are given by relations (\ref{elA}) and (\ref{magA}) for the Finslerian covariant derivative.

The electromagnetic field equations and the Bianchi identities for the $U(1)$ gauge field have been studied in Finsler geometry (see for example \cite{Miron} and for some recent papers \cite{Pfeifer:2011tk}). Keeping close to GR, the equations of motion for the electromagnetic field, with respect to the restricted absolute differentiation (\ref{tder}), are given by the simplified form
\begin{equation}
F^{ab}_{\;\;\;\;|b}=J^a.\label{Feqm}
\end{equation}
At first glance the equations of motion look the same with GR. However, a more careful examination of relation (\ref{Feqm}) reveals that 2nd order covariant differentiation of the vector potential is involved. For that reason, as in the Riemannian case in Sec.\ref{Rcove} the derivation of the electromagnetic wave equation requires the Finslerian Ricci identities. The latter, involve an extra contribution that directly originates from non-quadraticity or equivalently from the LV kinematics.

Taking into account the properties of the Cartan connection and assuming a vector field that depends only on the position coordinates, the Ricci identities take the simplified form
\begin{equation}
A^a_{\;\;|b|c}-A^{a}_{\;\;|c|b}=A^dR_{d\;\;bc}^{\;\;a}-C^a_{\;\;de}A^eR^{d}_{\;\;bc}\;,\label{Fric}
\end{equation}
where the $R$-torsion in the last term  is determined by the metric function and we can prove that it is related to the curvature tensor by the relation $R^{a}_{\;\;bc}=l^dR_{d\;\;bc}^{\;\;a}$. It is expected that LV are incompatible with the Ricci identities. Indeed, in Finsler geometry the effect of LV in the curvature
theory is given by the contribution of color in the rhs of relation (\ref{Fric}),  since Cartan's tensor (\ref{Cten}) is involved. On the other hand, using conservation arguments similar to GR we can relate the curvature tensor $R_{abcd}$ with the energy-momentum tensor (see for instance \cite{Miron,Kouretsis:2012ys}) by the following formula
\begin{equation}
R_{(ab)}-{1\over2}Rg_{ab}= T_{ab},\label{Ffield}
\end{equation}
where we use $8\pi G=1$ for the gravitational coupling constant, while $R_{ab}=g^{cd}R_{acbd}$ is the Ricci tensor and $R=g^{ab}R_{ab}$ is the Ricci scalar. The equations of motion for the electromagnetic field (\ref{Feqm}) combined with the modified Ricci identities (\ref{Fric}) and the algebraic relation between curvature and matter (\ref{Ffield}) determine the evolution of an electromagnetic field in our Finslerian setup.

Consider an irrotational and shear free bulk flow of matter. In that case, for the Finslerian covariant derivative we set ${\rm D}_a u_b={1\over3}\Theta h_{ab}$, in direct analogy to the FRW case discussed in Sec.(\ref{Rcove}). Then, on using the Coulomb gauge, decomposing the electromagnetic field equation (\ref{Feqm})  and taking its space-like part we arrive to the wave equation for the 3-vector potential given by relation (\ref{wavA}) for the Finslerian covariant derivative. The key difference is  that  the projected covariant derivatives appear in relation (\ref{wavA}) commute in a modified way due to the contribution of color in the Ricci identities (\ref{Fric}). In particular, projecting relation (\ref{Fric}) orthogonal to the observers' 4-velocity and using the Coulomb gauge we arrive to the contracted 3-Ricci identities
\begin{equation}
{\rm D}^{b}{\rm D}_{a}\mathcal{A}_b=\mathcal{R}_{ba}\mathcal{A}^b-\mathcal{G}_{ba}\mathcal{A}^b,\label{Ftric}
\end{equation}
where $\mathcal{R}_{ab}=h^{cd}\mathcal R_{acbd}$ is the Ricci curvature of the instantaneous rest frame given by relation (\ref{gaugo}) for the Finslerian covariant derivative, while $\mathcal{G}_{ba}=h_a^{\;\;c}h^{de}C_{dfb}R^f_{\;\;ce}$ is the projected effect of color. Using the Finslerian 3-Ricci identities (\ref{Ftric}) the wave equation (\ref{wavA}) is written as
\begin{equation}
\ddot{\mathcal{A}}_{\langle a\rangle} +\Theta \dot{\mathcal{A}}_{\langle a \rangle}+{1\over3} \left(\dot{\Theta}+\frac{2}{3}\Theta^2\right) \mathcal{A}_a-
{\rm D}^2\mathcal{A}_a+\mathcal{R}_{ba}\mathcal{A}^b-\mathcal{G}_{ba}\mathcal{A}^b=-\mathcal{J}_a,\label{FwavA}
\end{equation}
where the last term reflects the effect of color on the evolution of electromagnetic fields in our LV set-up. In other words, the adoption of a non-quadratic distance module breaks the Lorentz symmetry and as a result the Ricci identities are modified, directly affecting the evolution of the electromagnetic field.

 The spatial tensor $\mathcal G_{ab}$ given in relation (\ref{Ftric}) is a coupling term between the curvature and the Cartan tensor (\ref{Cten}). Therefore, the Finslerian contribution to the electromagnetic wave equation (\ref{FwavA}) is more likely to affect the evolution of seed fields in highly curved regions. Interestingly, that is the case  where we most likely expect effects of QG physics to emerge (e.g. the early universe).  To further investigate the evolution of magnetic fields in the Finslerian context we need to take a closer look to the coupling between curvature and LV appeared in relation (\ref{FwavA}). The curvature term $R_{abc}=l^dR_{dabc}$ follows the same symmetries with the Riemannian case \cite{RundFB} and for the isotropic and homogeneous limit is decomposed according to the following formula
\begin{equation}
 R_{abc}= {1\over3}Rg_{a[b}l_{c]}-R_{a[b}l_{c]}-g_{a[b}R_{c]d}l^d.\label{Rdec}
 \end{equation}
Then, on using the decomposition (\ref{Rdec}) together with the field equations (\ref{Ffield}) for a perfect fluid $T_{ab}=\rho u_a u_b+ph_{ab}$, and assuming a spatially flat geometry $\mathcal{R}_{ab}=0$, relation (\ref{FwavA}) takes the simplified form
\begin{equation}
\ddot{\mathcal{A}}_{\langle a\rangle} +\Theta \dot{\mathcal{A}}_{\langle a \rangle}+{1\over3} \left(\dot{\Theta}+\frac{2}{3}\Theta^2\right) \mathcal{A}_a-
{\rm D}^2\mathcal{A}_a-{1\over3}\rho\, \mathcal C_b \mathcal{A}^b \ell_a=-\mathcal{J}_a,\label{FwavA2}
\end{equation}
where $\mathcal C_a=h_a^{\;\;b}h^{cd}C_{bcd}$ is a spatial vector constructed by the Cartan tensor (\ref{Cten}) and $\ell_a=h_a^{\;\;b}l_b$ is the space-like part of the supporting direction. The last term in the lhs of relation (\ref{FwavA2}) monitors the anisotropic character of the theory in accordance to the LV phenomenology. As it is expected, the effect of breaking Lorentz invariance  depends on the orientation of the 3-vector potential in space \footnote{It is worth noting that since the model we discuss preserves the $U(1)$ symmetry the problematic ghost mode identified in \cite{Himmetoglu:2009qi} is absent. However, to ensure that in general no ghost modes exist  deserves further investigation.}. Moreover, in the particular limit we investigate, the Finslerian curvature theory unfolds a possible LV amplification mechanism that is more efficient in early phases of the cosmological evolution. Precisely,  the LV term in the wave equation is proportional to the energy density of matter. Hence, even for small departures from quadraticity the evolution of the electromagnetic field can be crucially affected in the first stages of the cosmological evolution. Interestingly, the high values of energy density  we expect in the early universe imply an electromagnetic field sensitive to departures from Lorentz invariance.

Concerning the evolution of the expansion in relation (\ref{FwavA2}), we expect that the modified Ricci identities will complicate the kinematics. For our case, where the supporting direction is parallel transported and the electromagnetic field is treated as a first order quantity,  the Raychaudhuri's equation \cite{Kouretsis:2012ys} for shear and vorticity free geodesics writes to
\begin{equation}
\dot \Theta+{1\over3}\Theta^2=-R_{ab}u^au^b-\mathfrak{T}_{ab}u^au^b,\label{Raych}
\end{equation}
where $\mathfrak{T}_{ab}u^au^b=C_{ac}^{\;\;\;\;d}l^eR_{e\;\;db}^{\;\;c}$ is the contribution of color to the expansion dynamics. The above relation, together with the wave equation (\ref{FwavA2}) for a particular form of Cartan's tensor (\ref{Cten}) determine the evolution of electromagnetic fields. In case of the shear and vorticity free kinematics that we imposed, and using the decomposition (\ref{Rdec}) together with the field equations (\ref{Ffield}), it is straightforward to prove that $\mathfrak{T}_{ab}u^au^b=0$ when the supporting direction is purely space-like, $l_au^a=0$. In that case, the expansion of the cosmic medium coalesces with the FRW universe  at the background level. However, on the perturbed manifold of the Finslerian medium the electromagnetic seed follows a different evolution history from the FRW case. In addition, we expect that the dynamo mechanism will also generate some shear and vorticity on the cosmic flow however their contribution is of second order in Raycaudhuri's equation \cite{Relcos,Kouretsis:2012ys}. The main suspect for the different dynamical behavior of the  electromagnetic perturbation is the modified 3-Ricci identities through which the  3-vector potential `feels' the color (in other words the LV) of space-time due to its vector nature.

\section{The amplification mechanism}

 In our set up,  the bulk flow of matter is given by the restricted Finslerian congruences along which the supporting direction is parallel transported, $\dot l^a=0$ \cite{RundFB,Kouretsis:2012ys}. In that case, if we assume that shear and vorticity are of first order and that $l_a$ is purely space-like, Raychaudhuri's formula (\ref{Raych}) implies an almost FRW expansion.  The supporting direction reflects the local anisotropy and  fits well with the LV character of the theory. In a LV framework we expect that the evolution of the electromagnetic field depends on the relative direction of the field and the preferred direction(s) induced by the broken symmetry. Within our Finslerian setup, the latter case is clearly depicted in the wave equation (\ref{FwavA2}) since  the evolution of the 3-vector potential depends on its orientation. From a theoretical viewpoint, the explicit-like violation induced by a constant supporting direction may not be completely satisfactory but we believe that our effective 'simplified' model catches the main characteristics of Lorentz symmetry breaking.

 As we already mentioned, the important point in our analysis is that the LV term in relation (\ref{FwavA}) is proportional to the energy density of matter. Apparently, in the early stages of the cosmic evolution, where the energy density is high,  the electromagnetic field is more sensitive to departures from Lorentz symmetry. Thus, we describe by geometric means the common belief of QG phenomenology that the higher the curvature of space-time, the stronger the effects of Lorentz symmetry breaking. The geometric entity that parameterizes LV is the Cartan tensor (\ref{Cten}) that measures the color (non-quadraticity) of space-time.

    Using the properties of Cartan's tensor (\ref{Cten}), the spatial vector $\mathcal C_a$ that enters the lhs of relation (\ref{FwavA2}) is always transverse to the purely spatial supporting direction, $\mathcal C_a\ell^a=0$. Given this orthogonality condition and assuming the harmonic decomposition $\mathcal A_{a}=\mathcal A_{(k)}\mathcal Q_a^{(k)}$ with $\dot {\mathcal Q}_a^{(k)}=0$ and  ${\rm D}^2\mathcal Q_a^{(k)}=-{k^2\over a^2}\mathcal Q_a^{(k)}$, the wave equation (\ref{FwavA2}) is written in the following form
   \begin{equation}
   \tilde{\mathcal{A}}_{(k)}''+a\sigma\tilde{\mathcal{A}}_{(k)}'+(k^2-{1\over3}a^2\rho\,\mathcal C)\tilde{\mathcal{A}}_{(k)}=0.\label{Fkwav}
   \end{equation}
where tilde stands for the rescaling $\tilde{\mathcal{A}}_{(k)}=a{\mathcal{A}}_{(k)}$, while we define $\mathcal C=\sqrt {\mathcal C_a\mathcal C^a}\tan(\theta)$ for the angle $\theta$ between the supporting direction and the 3-vector potential ($\theta\neq \pi/2$).   Also, we used Ohm's law $\mathcal J_a=\sigma E_a$ together with relation (\ref{elA}) for the Finslerian covariant derivative. The Laplace-Beltrami operator  within a Finslerian set-up is a well defined mathematical concept \cite{Antonelli}. The intrinsic torsion of the Finslerian manifold will affect the spectrum of the Laplace-Beltrami operator. However, for simplicity we use the standard harmonic decomposition since as we discussed after relation (\ref{Raych}) the background kinematics is `almost' FRW. In the following we will   use a mean field approximation (see for example \cite{Campanelli:2008xs}) as a first attempt to analyze the electromagnetic wave equation (\ref{Fkwav}). Also, notice that well within the horizon we recover the standard FRW wave equation and hence in an inflationary era we can interpret the electromagnetic seed as a `quantum fluctuation' as in the usual inflationary scenario. Nevertheless, due to the exponential expansion the coherence scale of the seed field grows rapidly and becomes a classical field.  The wave equation (\ref{Fkwav}) is a harmonic oscillator with varying parameters. In the extremal case  where the 3-vector potential is proportional or perpendicular to the supporting direction ($\theta=0$, $\theta=\pi/2$ respectively), the Finslerian contribution in the wave equation (\ref{Fkwav}) vanishes. The conductivity plays the role of damping while the LV contribution in the parenthesis implies a time varying frequency.

If the conductivity is negligible, the evolution of the electromagnetic field is determined by the varying frequency term in relation (\ref{Fkwav}). When the latter is negative we expect that for particular profiles of $\mathcal C$ the magnetic energy density is super-adiabatically amplified. This condition is fulfilled when the LV term dictates in the last term of relation (\ref{Fkwav}). For an almost flat FRW background, $H^2\sim{1\over3}\rho$, the required condition for super-adiabatic amplification reads
\begin{equation}
\mathcal{C}>\left(\frac{\lambda_H}{\lambda_{phys}}\right)^2,\label{Cineq}
\end{equation}
where we define $\lambda_H=1/H$ for the Hubble horizon and $\lambda_{phys}=a/k$ is the comoving wave-length of the electromagnetic fluctuation. Within the de Sitter inflationary scenario the coherence scale of the magnetic quantum fluctuation is well-outside of the horizon for the largest portion of its cosmological evolution, $\lambda_{phys}\gg \lambda_H$. Hence,  inequality (\ref{Cineq}) holds even for small values of $\mathcal{C}$. In other words electromagnetic fields of super-horizon scales are sensitive to departures from Lorentz invariance. The main reason for this effect is the color-curvature coupling in the last term of the Finslerian 3-Ricci identity (\ref{Ftric}). Due to this coupling the larger the coherence length, the more  the color affects the photon sector.

\subsection{Evolution of the  magnetic field}

The Cartan tensor $C_{abc}$ parameterizes LV since it monitors the colorful morphology of the Finslerian space-time. The particular form of the tensor $C_{abc}$ remains unconstrained unless we relate the metric function $F(x,y)$ with a particular QG scenario which is beyond  the scope of the present work. As an illustrative example let us  choose a  particular profile for the LV parameter $\mathcal{C}$. To be precise,   during inflation  we assume the simple case $\mathcal{C}\sim const$ in order to slow down the dilution of the magnetic intensity. Then, as we enter the classic era of radiation we set $\mathcal{C}\sim a^{-x}$ to `wash out' the QG effects. In a more detailed analysis one may study more complicated profiles for the Cartan's parameter $\mathcal{C}$ to avoid possible back-reaction and similar to strong coupling issues \cite{Demozzi:2009fu}.

In the inflationary era the temperature exponentially decays and the conductivity becomes negligible. Also, the scale factor evolves with respect to the conformal time as $a\propto\eta^{-1}$, with $\eta<0$. In addition, the Hubble parameter remains constant, $H=-(a_{INF}\eta_{INF})^{-1}$. Putting these together and using  the Friedmann equation $H^2\sim \frac{1}{3}\rho$ the wave formula of the 3-vector potential (\ref{Fkwav}) takes the simplified form
\begin{equation}
\tilde{\mathcal{A}}_{(k)}''+(k^2-\eta^{-2}\mathcal {C}_{INF})\tilde{\mathcal{A}}_{(k)}=0,\label{FDEinf}
\end{equation}
 which is of Bessel type and $\mathcal {C}_{INF}$ stands for the constant value of the LV parameter during de Sitter inflation. Since inflation implies very long coherence scales, after first horizon crossing we are interested in electromagnetic fields well outside the horizon, $k\eta\ll 1$. Then, the mean value of 3-vector potential for superhorizon modes evolves as $\tilde{\mathcal{A}}\propto \eta ^{\frac{1}{2}(1\pm \sqrt{1+4\mathcal{C}_{INF}}) }$. By virtue of relation (\ref{magA}) we get $\tilde{\mathcal{A}}\sim ka^2B$ for the mean values of the 3-vector and the magnetic intensity. Thus, in the inflationary era the generated  magnetic quantum fluctuation grows as $B\propto a^{-\nu/2}$ where $\nu=5\mp \sqrt{1+4\mathcal{C}_{INF}}$.  In this case, due to the approximate relation $B\sim k\eta E$ and observing that $\eta\propto a^{-1}$ the average intensity of the electric field scales as $E\propto a^{-\frac{1}{2}(\nu-2)}$. Thus, in order to ensure that  the energy density of the electromagnetic field remains smaller than the energy density of inflation, we get $\nu\geq2$. In that case the model avoids back-reaction issues \cite{Demozzi:2009fu}.

 After the end of inflation and assuming instantaneous reheating ($T_{RH}\sim M$) the cosmic fluid enters the radiation era. In this case, the scale factor evolves with respect to the conformal time as $a\propto \eta$. During radiation,   the conductivity is approximately proportional to temperature $\sigma\sim T/\alpha$, where $\alpha$ is the fine structure constant. Therefore, almost from the beginning of radiation the conductivity is very large  and at some point dominates in the wave equation (\ref{Fkwav}). Taking the limit of infinite conductivity $\sigma\gg H$, relation (\ref{Fkwav}) implies that $\mathcal{A}'\rightarrow0$ and on using (\ref{magA}) we recover the adiabatic decay $B\propto a^{-2}$. However, as we have already discussed in Sec.\ref{Rcove} one may argue that in superhorizon scales spatial currents are not important and the conductivity term in relation (\ref{Fkwav}) vanish (see for example \cite{Widrow:2011hs}). When the last term in relation (\ref{Fkwav}) dominates and assuming the power-law profile for the LV contribution $\mathcal {C}\propto a^{-x}$, the general solution for superhorizon modes ($k\eta\ll1$) reads
 \begin{equation}
 \tilde{\mathcal{A}}_{(k)}(\eta)=C_1\sqrt{\eta}\,{\rm K}_{{x/2}}(\zeta)+C_2\sqrt{\eta}\,{\rm I}_{{-x/2}}(\zeta),
 \end{equation}
for the modified Bessel functions with $\zeta={2\over x}\sqrt{\mathcal{C}}$. Therefore, at leading order the mean value of the 3-vector potential evolves as $\tilde{\mathcal{A}}\propto a$ which implies that the average value of the magnetic field grows superadiabatically, $B\propto a^{-1}$. This mechanism is valid until either the conductivity term in relation (\ref{Fkwav}) dominates or until the coherence length of the magnetic fluctuation approaches the horizon and condition (\ref{Cineq}) is no longer satisfied. After both these cases, the amplification stops operating and the magnetic energy density  follows the usual power law profile, $\rho_{B}\propto a^{-4}$.

\subsection{Residual magnetic field}

As we have already discussed, in standard relativistic electrodynamics magnetic fields dilute adiabatically. In the Finslerian setup the color directly implies the breaking of Lorentz invariance and under certain conditions can slow down the decay of the magnetic intensity. Roughly speaking, for an optimistic scenario  the energy density of the magnetic fluctuation is determined by the uncertainty principle, $\Delta\mathcal{E}\Delta t\sim1$. Taking into account that in the background the expansion dynamics are almost FRW (see relation (\ref{Raych})) we estimate the energy density at first horizon crossing , $(\rho_{B})_{HC}=\Delta \mathcal{E}/\Delta V\sim H^4$ \cite{Dimopoulos:2001as}. Then, using the power law for the magnetic intensity that we derived during the de Sitter era we get at the end of inflation $B_{INF}=B_{HC}(a_{HC}/a_{INF})^{\nu/2}$. After the end of inflation and for instantaneous reheating the cosmic fluid enters the epoch of radiation  where  plasma effects are important. We consider separately the two different possibilities, depending on whether or not spatial currents are important in superhorizon scales.

If plasma effects are negligible in scales well-outside the horizon, the LV amplification mechanism can last for a considerable amount of time in the radiation phase. The longer the scale of the fluctuation the more is affected by departures from the Riemannian measurement. However, during radiation the horizon grows faster than the physical wave-length  of the magnetic perturbation. Therefore, within a finite time-interval the parenthesis in relation (\ref{Fkwav}) becomes positive and we recover the adiabatic profile. Roughly speaking, this condition reads $\lambda_{phys}\sim(H^2\mathcal{C})^{-1/2}$, which corresponds to a particular value of the scale factor $a=a_*$. In fact,  taking into account that during radiation $aT\sim const$ and assuming a power-law profile for the LV-parameter $\mathcal{C}\propto a^{-x}$ we retrieve the approximate value for the scale factor
 \begin{equation}
a_*\sim \left (10^{-6} \frac{ \lambda_{0} }{ Mpc }\sqrt{\mathcal {C}_0 } \right)^{2/(2+x)},\label{Ast}
\end{equation}
 where $\mathcal{C}_0$ is the value of color today, $\lambda_0$ the scale of the B-field today, while the Hubble parameter is given with respect to the temperature by $H\sim T^2/m_{pl}$ and we used $T_0\simeq 2.35\times 10^{-13}GeV$ for the present temperature of the CMB. Notice that  the condition $\mathcal{C}_*<1$ implies that the adiabatic decay is recovered before the magnetic coherence scale re-enters the horizon. Thus, if we consider that the conductivity is negligible on superhorizon scales the plasma effects act after the adiabatic evolution of the magnetic field is recovered.

On the other hand, if we assume that plasma effects are important in super-horizon scales the amplification mechanism stops when the friction term in relation (\ref{Fkwav}) dominates. Approximating $a\sigma\tilde{\mathcal{A}}'$ with $a\sigma\tilde{\mathcal{A}}/\eta$ and using $a\eta\sim H^{-1}$ the conductivity dominates when ${\sigma/H}\gg\mathcal C$. The conductivity is approximately $\sigma\sim T/\alpha$ and given that $H\sim T^2/m_{pl}$,  on the onset of radiation we get ${\sigma/H}\sim{m_{pl}/ M\alpha}$ for instantaneous reheating. The upper value for the scale of inflation $M$ is constrained by the CMB for the spectrum of the  gravitational waves generated during inflation, $M\lesssim10^{-2}m_{pl}$. This translates to the lower limit for the conductivity after  instantaneous reheating ${\sigma/H}\gtrsim10^4$, hence for reasonable values of color any LV contribution is `washed-out'. Therefore, if we assume that conductivity is important on scales well-outside the horizon, the magnetic field evolves adiabatically from the end of inflation until today.

Finally, evolving the magnetic quantum fluctuation from the time of first horizon crossing until today, we get
\begin{equation}
B_0\sim 10^{-31}e^{-\nu N/2 }\frac{M^3}{m_{pl}}\left (10^{-6} \frac{ \lambda_{0} }{ Mpc }\sqrt{\mathcal {C}_0 } \right)^{ \frac{2}{2+x}(1-s) }\left(10^{32}\frac{M}{m_{pl}}\right)^{-s},\label{estmag}
\end{equation}
where $s$ is a `switch' that turns on and off the conductivity in superhorizon scales. When $s=0$ the plasma effects are omitted in superhorizon scales and the magnetic field decays adiabatically after $a_{*}$ given in (\ref{Ast}), while for $s=1$ any LV effect is subdominant immediately after inflation. Apparently, for a wide range of the parameters of inflation and Lorentz symmetry violation, we can sustain astrophysically relevant magnetic fields  with coherence scales  $10kpc-1Mpc$. Interestingly, taking into account the observational constraints on primordial magnetic fields, relation (\ref{estmag}) gives back constraints for the `Finslerity' of space-time. Indeed, there are various constraints on primordial magnetism from matter density fluctuations \cite{Yamazaki:2008jh}, the laser interferometer gravitational-wave observatory (LIGO) \cite{Wang:2008vp}, the Chandra X-ray and Sunyaev-Zel'divich surveys \cite{2012MNRAS.424..927T} and from the Big Bang Nucleosynthesis \cite{grasso}. All the previous investigations provide weaker constraints than the CMB which gives back an upper limit of $\sim10^{-9}G$ \cite{CMB} that can be potentially upgraded up to $\sim10^{-11}G$ by  CMB polarization experiments \cite{CMBupg}. From the latter constraints, assuming that the energy density of the electromagnetic field decreases during inflation $(\nu=2)$, for a coherent seed field of $1Mpc$  and the upper value of the inflationary scale $M=10^{-2}m_{pl}$ we get the range for the today values of the Cartan's parameter, $10^{-53}\lesssim\mathcal{C}_0\lesssim10^{-6}$. Also, there exist constraints that practically rule out a blue spectrum of the inflation generated magnetic field \cite{Caprini:2001nb}. The primordial magnetic field generates gravitational waves and constraints from Nucleosynthesis point towards a red spectrum. We expect that the properties of the spectrum within our setup will be affected by the colored 'morphology' of the Finslerian space-time. This detailed and complicated analysis will eventually constrain further the parameter space of the LV setup and we will address it to future work.

 In addition, we can recover standard electromagnetism at BBN and still sustain relevant astrophysical magnetic fields. Given that $T\sim1MeV$ during BBN relation (\ref{Ast}) gives back the upper value for the `Finslerity' today $\mathcal{C}_0\lesssim10^{-12}$ that translates to a magnetic field $B_0\lesssim10^{-13}G$ on $1Mpc$.  Putting all these together, the 'safe' range of the Cartan's parameter is
 \begin{equation}
 10^{-53}\lesssim\mathcal{C}_0\lesssim10^{-12}\label{Conol}.
 \end{equation}
The parameter $\mathcal C$ represents a potential Lorentz invariance violation in the gravity sector. In the absence of  curvature we recover the standard electromagnetic theory. However, when the coherence scale is larger than the horizon curvature becomes important and  the primordial seed becomes sensitive to possible LV in gravitational physics. The range of the Cartan's parameter (\ref{Conol}) results exactly from this mechanism. Therefore, constraints that arise from LV in the gravitational sector can be potentially combined with the effect of color on primordial magnetic fields that leads to the estimate (\ref{Conol}) for the 'Finslerity' of space-time. The reader should note that, the parameterized post-Newtonian analysis in Finsler geometry has been studied in  \cite{PPNfin}.

Furthermore, if we take into account spatial currents in superhorizon scales after inflation the model suffers from backreaction/strong coupling issues \cite{Demozzi:2009fu}. For the best case scenario of a decaying electromagnetic energy density ($\nu=2$) and for the extremal value for the scale of inflation $M=10^{-2}m_{pl}$ we can sustain a primordial magnetic field  of $10^{-33}G$ on $1Mpc$. This magnetic intensity can be astrophysically relevant only for the most optimistic scenarios of the galactic dynamo and protogalactic collapse. However, in a more sophisticated approach we may consider non-monotonic profiles of the Cartan's parameter $\mathcal{C}$, to overcome the backreaction/strong coupling problem \cite{Demozzi:2009fu}.

\section{Discussion}

Summarizing, in this article we concentrate our analysis on the possibility of sustaining seed magnetic fields within an inflationary scenario. Motivated from current studies of QG theories,  the amplification mechanism  is based on the assumption that Lorentz invariance is not an exact symmetry of nature.   In our analysis, we try to formulate departures from Lorentz invariance with pure geometric means. To achieve this, we first develop the 4-potential representation of electromagnetism in a covariant framework by following the 1+3 formalism. The derivation of the wave equation for the 3-vector potential brings in the center of attention the Ricci identities. The latter identities, are incompatible with a large class of LV theories \cite{Kostelecky:2003fs} and this creates an obstacle to investigations of electrodynamics in curved space-times when we abandon Lorentz symmetry.  From a geometric perspective we expect that the evolution of the electromagnetic field will be affected through some modified commutation formula for vector fields. This discussion is in alliance with amplification mechanisms that break the conformal symmetry of electromagnetism, since Lorentz invariance underpins the local structure of GR as well as the symmetries of the electromagnetic field theory.

 We use Finsler geometry as our theoretical background to study the evolution of electromagnetic fields when Lorentz invariance is broken.  Finsler geometry provides a general framework to study LV theories since we relax the local symmetries of the space-time manifold by dropping the quadratic restriction on the distance module. The LV is parameterized by the Cartan's tensor that measures departures from the Riemannian measurement and enters the generalised Ricci identities. Keeping close to GR, we derived a modified wave equation  for the electromagnetic field. The LV contribution directly originates from the Finslerian Ricci identities that monitor the curved and locally anisotropic structure. Interestingly, the geometric amplification is based on a coupling term in the wave equation, between the curvature of space-time and the Cartan tensor that parameterizes LV. This introduces an amplification mechanism that is more efficient in long wavelengths where curvature effects are important. Thus, within an inflationary scenario where quantum fluctuations are stretched to superhorizon scales, the Finslerian amplification mechanism can generate sufficiently strong magnetic fields in coherence scales $10kpc-1Mpc$.

 The introduction of color breaks the local symmetries of GR and modifies the electromagnetic field theory. Within our setup, the color of space-time is measured by the Cartan tensor that evolves throughout the cosmological history. Due to its presence, we will get additional LV effects on other cosmological relics. For example, the non-quadratic metric function implies a modified mass-shell condition \cite{Rad41} that can potentially be constrained by the baryon asymmetry and the abundance of light elements. Also, the present  amplification mechanism will induce some anisotropisation of the CMB background. Moreover, we can estimate the upper bounds of the parameters that break the quadratic restriction by collider physics, threshold anomalies, time of flight effects and solar system tests. All this phenomenology and cosmological constraints combined with particular QG theories where the effective geometry is Finslerian, can give a better insight about LV and the colorful morphology of the physical manifold. This detailed and complicated analysis is an intriguing open challenge (for some recent studies see \cite{Mavromatos:2010ar,Lammerzahl:2012kw}). Since our amplification mechanism strongly depends on the intensity of Cartan tensor and its evolution in time,  constraining the `Finslerity' of space-time may lead to a better understanding of why our universe is magnetised.

\section{Acknowledgements}
The author wishes to thank Christos Tsagas for introducing him to the topic and for his encouragement. The author also thanks Michalis Stathakopoulos and Panagiotis Stavrinos for useful discussions and comments.

\end{document}